\pdfoutput=1

\documentclass[preprint,10pt]{elsarticle}
\usepackage{hyperref}

\makeatletter
\def\ps@pprintTitle{%
    }
\makeatother

\usepackage{amssymb}
\usepackage{amsmath}
\usepackage{mathtools}

\begin{document}

\begin{frontmatter}

\title{High-order implicit solver in conservative formulation for tokamak plasma transport equations}

\author{Andrei Ludvig-Osipov\corref{cor1}}
\ead{osipov@chalmers.se}
\cortext[cor1]{Corresponding author}
\author{Dmytro Yadykin}
\author{Pär Strand}
\address{Department of Space, Earth and Environment, Chalmers University of Technology, SE-412~96, Gothenburg, Sweden\vspace{-0.2cm}}

\begin{abstract}
%% Text of abstract

An efficient numerical scheme for solving transport equations for tokamak plasmas within an integrated modelling framework is presented.
The plasma transport equations are formulated as diffusion-advection equations in two coordinates (a temporal and a spatial) featuring stiff non-linearities.
The presented numerical scheme aims to minimise computational costs, which are associated with repeated calls of numerically expensive physical models in a processes of time stepping and non-linear convergence within an integrated modelling framework.
The spatial discretisation is based on the 4th order accurate Interpolated Differential Operator in Conservative Formulation, the time-stepping method is the 2nd order accurate implicit Runge-Kutta scheme, and an under-relaxed Picard iteration is used for accelerating non-linear convergence.
Temporal and spatial accuracies of the scheme allow for coarse grids, and the implicit time-stepping method together with the non-linear convergence approach contributes to robust and fast non-linear convergence.
The spatial discretisation method enforces conservation in spatial coordinate up to the machine precision.
The numerical scheme demonstrates accurate, stable and fast non-linear convergence in numerical tests using analytical stiff transport model.
In particular, the 2nd order accuracy in time stepping significantly improves the overall convergence properties and the accuracy of simulating transient processes in comparison to the 1st order schemes.

\end{abstract}


\end{frontmatter}

%% \linenumbers

%% main text
\section{Introduction}
\label{sec:intro}

Numerical simulations of magnetically confined plasmas are crucial for the design, development and exploitation of nuclear fusion experiments.
The integrated (multi-physics) modelling 
has been extensively used in analysis and modelling campaigns for existing tokamaks (e.g., JET~\cite{Kim+etal2023}, ASDEX Upgrade~\cite{Luda+etal2020,Fajardo+etal2024}, JT60-SA~\cite{Ostuni+etal2021}) to pave way for the future experiments -- ITER~\cite{Mailloux+etal2022overview,Garcia+etal2019first} and DEMO~\cite{Litaudon+etal2022}.

In integrated modelling schemes~\cite{Coster+etal2010,Cenacchi+Taroni1988jetto,Pereverzev+Yushmanov2002,Romanelli+etal2014jintrac,Hayashi+etal2007,Honda+Fukuyama2008task,Hawryluk1981transp,Artaud+etal2010} the plasma transport equations are solved to evolve macroscopic quantities such as temperatures and densities of particle species, and plasma's electric current and rotation.
The transport equations form a self-consistent model of tokamak plasma in terms of macroscopic quantities, averaged over the closed nested magnetic flux surfaces (topologically toroidal surfaces 
enclosing constant magnetic flux).
The physical models for equilibrium, transport, and heating and current drive, which represent the geometry, diffusivities and source terms respectively, provide the coefficients to the transport equations.
For our purposes we assume the magnetic equilibrium to be static~\cite{Grad+Hogan1970,Blum+LeFoll1984plasma}.
The resulting transport equations form a system of stiff nonlinear coupled partial differential equations in two coordinates -- the time and the flux surface label.
The stiffness to large extent comes from the turbulent transport, dominating in typical tokamak scenarios -- the turbulent particle and energy fluxes sharply increase when the plasma density and temperature gradients surpass certain critical values.

The choice of a turbulent transport model for the integrated modelling simulations is a trade-off between the degree of fidelity and the computational costs.
The fully non-linear gyrokinetic models, such as GENE~\cite{Jenko+Dorland2001} and CGYRO~\cite{Candy+etal2016} offer high physical accuracy at a price of high numerical costs.
The turbulent transport codes, more commonly used in integrated modelling frameworks, are the quasilinear models, such as TGLF~\cite{Staebler+etal2005}, QLK~\cite{Citrin+etal2017} and EDWM~\cite{Strand+etal2004,Tegnered+etal2016,Fransson+etal2022}, and semi-impirical models, such as Bohm-gyroBohm~\cite{Erba+etal1997} and CDBM~\cite{Fukuyama+etal1998}.
To achieve robust simulation results in an integrated modelling scheme, the non-linearities in plasma transport equations should be resolved at each time step by an iterative algorithm.
This in general implies calls to a transport model and other physical modules at every iteration.
An efficient numerical scheme should be used to minimise the total number of iterations and the numerical costs of each individual iteration, and therefore a computational cost of an entire simulation with the goal of enabling a routine usage of higher fidelity models.

The total number of iterations can be reduced by the efficiency of a non-linear convergence scheme through accelerated resolution of non-linearities, and by improvement of the accuracy with respect to the temporal grid through allowing longer time steps.
The numerical cost of the majority of the physical models, used as components in integrated modelling schemes, scales with the resolution in spatial (flux surface) coordinate, hence a robust convergence of a numerical scheme in spatial grid size can reduce the required computational resources for individual iterations.
An essential aspect of a solver for plasma transport equations is its conservation properties. 
The transport equations are conservation laws for particles, energy, magnetic flux and momentum; therefore the numerical scheme must be consistently conservative to avoid errors in conserved quantities, which can accumulate during temporal evolution.

The numerical schemes in integrated modelling frameworks, presently used in fusion research, employ various combinations of spatial discretisation, time-stepping and non-linear convergence methods.
Among the spatial discretisation schemes are finite difference in ETS~\cite{Coster+etal2010}, JETTO~\cite{Cenacchi+Taroni1988jetto} and TRANSP~\cite{TRANSP_code}, finite element in RAPTOR~\cite{Felici+etal2018} and TASK/TX~\cite{Honda+Fukuyama2008task}, and interpolated differential operation in FASTRAN~\cite{Park+etal2017}.
The most commonly used time stepping method is backward Euler (the first order implicit scheme) employed in ETS~\cite{Coster+etal2010}, FASTRAN~\cite{Park+etal2017}, TASK/TX~\cite{Honda+Fukuyama2008task} and RAPTOR~\cite{Felici+etal2018}, and often combined with adaptive time step size;  and the two-point difference derivative approximation is combined with predictor-corrector in JETTO~\cite{Cenacchi+Taroni1988jetto} for time-advance and resolution of non-linearities.
To accelerate the convergence of stiff non-linearities, the Newton iterations are used in RAPTOR~\cite{Felici+etal2018} and TRANSP~\cite{TRANSP_code}, artificial diffusivity with compensation by advective or source terms~\cite{Pereverzev+Corrigan2008} in ETS~\cite{Coster+etal2010}, ASTRA~\cite{Pereverzev+Yushmanov2002} and JETTO~\cite{Cenacchi+Taroni1988jetto}, and root-finding algorithms (under-relaxed Picard iteration and secant method) in FASTRAN~\cite{Park+etal2017}.

In this work, a solver for transport equations is proposed, which aims to minimise the number of calls to physical models, is intrinsically conservative and accurate, and demonstrates stable non-linear convergence for a problem with stiff transport.
The spatial discretisation is based on the 4th order Interpolated Differential Operation in Conservative Formulation (IDO-CF)~\cite{Imai+etal2008}, for the time stepping the 2nd order implicit Runge-Kutta scheme is used~\cite{Wanner+Hairer1996solving}, and the non-linear convergence is accelerated by the under-relaxed Picard iteration~\cite{Linge+etal2017}.
The main advantage of the proposed solver is the 2nd order of convergence in time, which enables accurate simulation of transient processes on coarse temporal grids.

The paper is organised as follows.
Sec.~\ref{sec:method} introduces the general form of a transport equation, describes the conservative numerical scheme for a linearised iteration and the nonlinear convergence algorithm.
The results demonstrating the performance of the proposed numerical scheme are reported in Sec.~\ref{sec:results}.
Sec.~\ref{sec:conclusions} summarises and concludes the paper.

\section{Numerical scheme}
\label{sec:method}

We consider the energy equation in axisymmetric toroidal geometry~\cite{Hinton+Hazeltine1976} as an example of a one-dimensional transport equation 
\begin{equation}
    \frac{3}{2}\frac{\partial (V' n T)}{\partial t} +
    \frac{\partial }{\partial \rho}
    \left[
        V' 
        \left(
            - \left\langle |\nabla \rho|^2 \right\rangle  
            \chi \, n \frac{\partial T}{\partial \rho} 
            + \left\langle |\nabla \rho| \right\rangle
            V^\mathrm{pinch} n T
        \right)
    \right]
    =
    V'S,
    \label{eq:energy}
\end{equation}
where $n$ and $T$ are the flux-surface-averaged particle density and temperature, $\chi$ and $V^\mathrm{pinch}$ are the diffusivity and the pinch velocity, $S$ represents sources, $\rho$ -- radial coordinate (toroidal flux surface label) and $V'$ is a derivative of a volume inside a flux surface with respect to $\rho$.
The diffusivity $\chi$ is a highly nonlinear function of the temperature gradient $\frac{\partial T}{\partial \rho}$.
Integration of Eq.~\eqref{eq:energy} over $\rho$ shows the conservation property of a transport equation: the rate of change of energy $\int_{\rho_1}^{\rho_2}\frac{\partial(V'n T)}{\partial t}\mathrm{d}\rho$ between the two flux surfaces $\rho_1$ and $\rho_2$ is balanced by the total energy flux at these surfaces and the energy source in between these surfaces.
Similarly, the transport equations for the density, for the plasma current, and for the toroidal rotation are conserving the particles, the poloidal flux, and the toroidal angular momentum respectively~\cite{Hinton+Hazeltine1976}. 

The transport equations can be reformulated to a generalised form~\cite{Coster+etal2010} (the same for all the transport equations) which provides a unified interface to the numerical scheme:
    \begin{equation}
    \begin{split}
        \frac{\partial\left[ Y(x,t)\right]}{\partial t} = -
        \frac{\partial}{\partial x}\left( 
        -d(x,t) \frac{\partial Y(x,t)}{\partial x} +
        e(x,t)Y(x,t) \right) +
        f(x,t),
     \end{split}
     \label{eq:GFTE_dt1}
    \end{equation}
where $x$ is the normalized radial coordinate, $Y(x,t)$ is the conserved quantity, $d(x,t)$ and $e(x,t)$ are the effective diffusion and advection respectively, and $f(x,t)$ is the source term.
The boundary conditions at the ends of the radial interval $[x_{1}, x_{N}]$ are given as
    \begin{equation}
        u_{i}(t) Y(x_{i},t) + v_{i}(t) \frac{\partial Y(x,t)}{\partial x}\big|_{x=x_{i}}=w_{i}(t), i=1,N.
        \label{eq:bnd}
    \end{equation}
    Here $u_{i},v_{i}\neq0$ gives the Robin (mixed) boundary condition; $u_{i}=0,v_{i}\neq0$ gives the Neumann boundary condition; and $u_{i}\neq 0,v_{i}=0$ -- the Dirichlet boundary condition.

In the example for the energy equation~\eqref{eq:energy} we have $Y=\frac{3}{2}V'nT$, $d=\frac{2}{3}\left\langle |\nabla \rho|^2 \right\rangle \chi$, $e=\frac{2}{3}\left\langle |\nabla \rho| \right\rangle V^\mathrm{pinch} - \frac{2\left\langle |\nabla \rho|^2 \right\rangle (V'n)' \chi}{3V'n}$, and $f=V'S$.

The rest of this section presents a numerical scheme which is designed to solve Eqs.~\eqref{eq:GFTE_dt1}-\eqref{eq:bnd} for the conserved quantity $Y(x,t)$.
The numerical scheme operates as follows: at each time step, the nonlinearities are resolved with an iterative scheme, where at each iteration a linearised equation (with fixed transport equation coefficients $d(x),e(x),f(x)$) is solved.

The time advance scheme is discussed in Sec.~\ref{sec:time_advance}.
Sec.~\ref{sec:spatial} presents the spatial discretisation scheme.
In Sec.~\ref{sec:matrix} a numerical scheme for a linearized iteration with fixed transport coefficients is presented.
The convergence method to resolve the nonlinear dependence in transport coefficients on the spatial derivative of the profile $Y(x)$ is presented in Sec.~\ref{sec:nonlin_converg}, where at each iteration the linearized step is solved.

\subsection{Time advance}
\label{sec:time_advance}

The time integration is performed using a two-stage implicit Runge-Kutta method~\cite{Wanner+Hairer1996solving} with Lobatto $\mathrm{III_C}$ parameters~\cite{Wanner+Hairer1996solving,Ehle1969pade}.
This time integration method is both L-stable and algebraically stable, and thus suitable for stiff problems~\cite{Jay2015}.
The implicit Runge-Kutta methods are designed to solve equations of the form
\begin{equation}
    \frac{\mathrm{d} Y(t)}{\mathrm{d} t} = F_t[Y],
    \label{eq:iRK}
\end{equation}
where $F_t[Y]$ is a time-dependent operator that acts on the unknown $Y(x)$ and returns a function of $x$.
For the purposes of the present numerical scheme, the operator $F$ is represented by the right hand side of Eq.~\eqref{eq:GFTE_dt1}.
The dependence on the radial coordinate $x$ is implied, but not written out explicitly.

The two-stage Lobatto $\mathrm{III_C}$ method for the time stepping proceeds as follows:
\begin{align}
    \label{eq:lobatto1}
    Y(t) &= Y(t-\tau) + 0.5\tau (s_1 + s_2); \\
    \label{eq:lobatto2}
    s_1 &= F_{t-\tau} [Y(t-\tau) + 0.5\tau (s_1 - s_2) ]; \\
    \label{eq:lobatto3}
    s_2 &= F_t [Y(t-\tau) + 0.5\tau (s_1 + s_2)],
\end{align}
where $s_1,s_2$ are stages (intermediate variables), and $\tau$ is the time step.
The two-stage Lobatto $\mathrm{III_C}$ method is second-order accurate in time~\cite{Wanner+Hairer1996solving}.

As a reference, we use the backward Euler time advance scheme, which is first-order accurate in time and approximates Eq.~\eqref{eq:GFTE_dt1} as
\begin{equation}
    \frac{Y(x,t) - Y(x,t-\tau)}{\tau} = F_t[Y(x,t)](x),
    \label{eq:backwards_euler}
\end{equation}
i.e., the right hand side is taken at the current (predicted) time $t$ for the nonlinear stability reasons~\cite{Wanner+Hairer1996solving}.

\subsection{Spatial discretisation}
\label{sec:spatial}

In this section, we consider the spatial discretisation for a linearised iteration, that is, when the transport equation coefficients $d(x),e(x),f(x)$ are fixed (independent of $Y'$).
Then the operator $F_t[Y](x)$ given by the right hand side of Eq.~\eqref{eq:GFTE_dt1} can be represented as
\begin{equation}
    F_t[Y](x) = L_t[Y](x) + f(x,t),
\end{equation}
where $f(x,t)$ is the previously introduced source term function, and
\begin{equation}
    L_t[y](x):= (d(x,t)y'(x) - e(x,t)y(x))'
    \label{eq:Loperator}
\end{equation}
is a linear differential operator acting on some function $y(x)$.

To obtain the spatial discretisation of the linear operator $L_t[y](x)$ on the grid $\mathbf{x}=[x_i]_{i=1}^N$, the function $y(x)$ is approximated by a set of fourth degree polynomials.
In the following, we will use the shorthand $y_i:=y(x_i)$, $\tilde{y}_i:=\int_{x_i}^{x_{i+1}} y(x)\mathrm{d}x$.
The function $y$ at each two-cell interval $x_{i-1}<x<x_{i+1}$ ($i=2,..,N-1$) is approximated by a fourth degree polynomial
\begin{equation}
        P_i(\delta ) = k_4^{(i)}\delta^4 + k_3^{(i)}\delta^3 + k_2^{(i)}\delta^2 + k_1^{(i)} \delta + y_{i},
        \label{eq:polynomial}
\end{equation}
where $\delta = x-x_i$.
The polynomial coefficients $k_1^{(i)},k_2^{(i)},k_3^{(i)},k_4^{(i)}$ for the $i$th polynomial are obtained from the matching conditions
    \begin{align}
        P_i(x_{i+1}-x_{i}) &= y_{i+1}, 
        \label{eq:matching1} \\
        P_i(x_{i-1}-x_{i}) &= y_{i-1}, \\
        \int_{x_{i-1} - x_{i}}^0 P_i(\delta)\mathrm{d}\delta &= \tilde{y}_{i-1}, \\ 
        \int_0^{x_{i+1} - x_{i}} P_i(\delta)\mathrm{d}\delta &= \tilde{y}_{i},
        \label{eq:matching4}
    \end{align}
and are each a linear combination of $\{y_{i-1},y_i,y_{i+1},\tilde{y}_{i-1},\tilde{y}_{i}\}$, see~\ref{sec:polycoeff}.
The piece-wise polynomial approximation of $y(x)$ is, via the coefficients of the polynomials $P_{i}, i=2,..,(N-1)$, fully determined by a vector $\mathbf{y}=[y_1,..y_N,\tilde{y}_1,..,\tilde{y}_{N-1}]^\mathrm{T}$.
The spatial derivatives of the profile on the grid are obtained as $y'_i = P'_i(\delta)\big|_{\delta=0} = k_1^{(i)}$ and $y''_i = P''_i(\delta)\big|_{\delta=0} = 2 k_2^{(i)}$.

The differential operator~\eqref{eq:Loperator} is a linear combination of $y,y',y''$ and therefore can be approximated on the grid $\mathbf{x}$ as follows for $i=2,..,N-1$
\begin{equation}
    L_t[y](x_i) 
        = d(x_i,t)y''_i + (d'(x_i,t) - e(x_i,t))y'_i - e'(x_i,t)y_i
        \approx \pmb{\ell}_i \mathbf{y},
        \label{eq:ell_i}
\end{equation}
where $\pmb{\ell}_i$ is a row vector of size $2N-1$ with non-zero entries corresponding to the positions of $\{y_{i-1},y_i,y_{i+1},\tilde{y}_{i-1},\tilde{y}_{i}\}$ in $\mathbf{y}$.
In matrix notation,
\begin{equation}
    L_t[y](\mathbf{x}) \approx \mathbf{L}_t \mathbf{y},
\end{equation}
where $\mathbf{L}_t$ is an $(N-2)\times(2N-1)$ matrix constructed by stacking vectors $\pmb{\ell}_i$.
The matrix $\mathbf{L}_t$ depends only on the grid $\mathbf{x}$, the functions $d(x,t)$ and $e(x,t)$ with their spatial derivatives. The derivatives $d'(x,t), e'(x,t)$ are calculated from a cubic splines representation to retain a high order of spatial convergence of the scheme.

To solve the time-advance scheme, Eqs.~\eqref{eq:lobatto1}-\eqref{eq:lobatto3}, extra equations should be introduced to account for the additional free variables $\tilde{y}_1,..,\tilde{y}_{N-1}$.
To address that and to enforce conservation of cell-integrated quantities, we discretise the cell integrals of the operator $L$:
\begin{equation}
    \tilde{L}[y]_i :=
    \int_{x_i}^{x_{i+1}} L_t[y](x) \mathrm{d}x=
        d(x,t)y'(x) - e(x,t)y(x) \approx \tilde{\pmb{\ell}}_i \mathbf{y}
        \label{eq:tilde_ell_i}
\end{equation}
for $i=1,..,N-1$, where $\tilde{\pmb{\ell}}_i$ is a $2N-1$ row vector with non-zero entries at the same positions as in $\pmb{\ell}_i$.
Similarly, $\tilde{\pmb{\ell}}_i$ are stacked into a $(N-1)\times(2N-1)$ matrix $\tilde{\mathbf{L}}_t$ to get
\begin{equation}
    [\tilde{L}[y]_1,..,\tilde{L}[y]_{N-1}]^\mathrm{T} \approx \tilde{\mathbf{L}}_t\mathbf{y}.
\end{equation}

The left hand side of the boundary condition~\eqref{eq:bnd} can be represented as a scalar product
\begin{equation}
    u_i(t)y_i + v_i(t)y'_i \approx \mathbf{u}_i(t)\mathbf{y}, \quad i=1,N,
    \label{eq:u_l}
\end{equation}
where $\mathbf{u}_i(t)$ is a row vector of size $2N-1$.

The expressions for $\pmb{\ell}_i, \tilde{\pmb{\ell}}_i, \textbf{u}_i$ are given in \ref{sec:linear_transf}.
This spatial discretisation scheme is based on the spatial discretisation approach in Interpolated Differential Operator in Conservative Form presented in~\cite{Imai+etal2008}.
In the following, the here presented spatial discretisation scheme is referred to as IDO-CF.

\subsection{Matrix formulation}
\label{sec:matrix}

In this section a matrix equation for computing the time advance step~\eqref{eq:lobatto1}-\eqref{eq:lobatto3} with the boundary conditions~\eqref{eq:bnd} is presented.
For convenience we define a $(2N-1)\times(2N-1)$ matrix 
\begin{equation}
    \mathcal{L}_t = 
    \begin{bmatrix}
        \mathbf{L}_t \\
        \tilde{\mathbf{L}}_t \\
        \mathbf{u}_1(t) \\
        \mathbf{u}_N(t) 
    \end{bmatrix}.
\end{equation}
Then the time advance step~\eqref{eq:lobatto2}-\eqref{eq:lobatto3} is given by
\begin{equation}
    \begin{bmatrix}
        \mathbf{s}_1 \\
        \mathbf{s}_2
    \end{bmatrix}
    =
    0.5\tau
    \begin{bmatrix}
        \mathcal{L}_{t-\tau} & -\mathcal{L}_{t-\tau} \\
        \mathcal{L}_t        & \mathcal{L}_t
    \end{bmatrix}
    \begin{bmatrix}
        \mathbf{s}_1 \\
        \mathbf{s}_2
    \end{bmatrix}
    +
    \begin{bmatrix}
        \mathcal{L}_{t-\tau} \\
        \mathcal{L}_t
    \end{bmatrix}
    \mathbf{Y}(t-\tau)
    +
    \begin{bmatrix}
        \mathbf{f}(t-\tau) \\
        w_1(t-\tau) \\
        w_N(t-\tau) \\ 
        \mathbf{f}(t) \\
        w_1(t) \\
        w_N(t) 
    \end{bmatrix},
    \label{eq:matrix}
\end{equation}
where
$\mathbf{s}_j=[s_{j,1},..,s_{j,N},\tilde{s}_{j,1},..,\tilde{s}_{j,N-1}]^\mathrm{T}$ for $j=1,2$, $\mathbf{Y}=[Y_1,..,Y_N,\tilde{Y}_1,..\tilde{Y}_{N-1}]^\mathrm{T}$, and \quad $\mathbf{f}=[f_2,f_3,..,f_{N-1},\tilde{f}_1,..\tilde{f}_{N-1}]^\mathrm{T}$.
Eq.~\eqref{eq:matrix} represents a $(4N-2)\times(4N-2)$ linear system of equations and for ease of notation in the following subsection is written as
\begin{equation}
    \mathbf{A}
    \begin{bmatrix}
        \mathbf{s}_1 \\
        \mathbf{s}_2
    \end{bmatrix}
    =
    \mathbf{B}.
\end{equation}
The discretised solution for a profile at time $t$ is, in accordance with Eq.~\eqref{eq:lobatto1}, 
\begin{equation}
    \mathbf{Y}(t) = \mathbf{Y}(t-\tau) + 0.5\tau (\mathbf{s}_1 + \mathbf{s}_2).
\end{equation}

Additional coefficient functions can be straightforwardly introduced to Eq.~\eqref{eq:GFTE_dt1} with the appropriate modifications to the numerical scheme~\eqref{eq:matrix}.
For example, it is often useful to introduce a coefficient function in front of the time derivative to compensate for the numerical issues related to $V'\sim x$ close to $x=0$.
In this case, the left hand side of Eqs.~\eqref{eq:lobatto2} and~\eqref{eq:lobatto3} is modified to $a(x)s_j(x)$, $j=1,2$.
This, in turn, will require a discretisation to evaluate integrals of the form $\int_{x_i}^{x_{i+1}}a(x)s_j(x) \mathrm{d}x$, which can be performed with the Gaussian quadrature using the polynomial representation to evaluate $s_j$ at the quadrature points.
This will entail a corresponding matrix to multiply the left hand side of~\eqref{eq:matrix}.

Similarly the backward Euler time stepping~\eqref{eq:backwards_euler} scheme can be represented as
\begin{equation}
    \mathbf{Y}(t) = \mathbf{Y}(t-\tau) + \tau \mathcal{L}_t\mathbf{Y}(t) + 
    \begin{bmatrix}
        \mathbf{f}(t) \\
        w_1(t) \\
        w_N(t) 
    \end{bmatrix}.
\end{equation}

\subsection{Nonlinear convergence method}
\label{sec:nonlin_converg}

The transport coefficients, in particular the diffusivity $\chi$, are strongly dependent on the profile derivative $Y'(x)$.
Therefore, the matrix $\mathbf{A}$ is dependent on the profile vector $\mathbf{Y}(t)$ and the nonlinear equation
\begin{equation}
    \mathbf{A}(\mathbf{Y}(t))
        \begin{bmatrix}
            \mathbf{s}_1 \\
            \mathbf{s}_2
        \end{bmatrix}
    = \mathbf{B}
    \label{eq:matrix_nonlin}
\end{equation}
is to be solved at each time step.
Picard type operation with under-relaxation~\cite{Park+etal2017,Linge+etal2017} is used given an initial profile vector $\mathbf{Y}^{(0)}$ (a profile at the previous time step).
We denote the iteration index as $k$, and starting from the first iteration $k=1$ follow the steps:
\begin{enumerate}
    \item At iteration $k$, the matrix is approximated as
    \begin{equation}\nonumber
        \mathbf{A}^{(k)} = 
    \begin{cases}
        \mathbf{A}(\mathbf{Y}^{(k-1)}), &  k=1; \\
        \alpha \mathbf{A}(\mathbf{Y}^{(k-1)}) + (1-\alpha)\mathbf{A}^{(k-1)}, & k>1.
    \end{cases}
    \end{equation}
    \item The approximation of the profile at $k$th iteration $\mathbf{Y}^{(k)}$ is obtained from the solution of $\mathbf{A}^{(k)}
    \begin{bmatrix}
        \mathbf{s}_1 \\
        \mathbf{s}_2
    \end{bmatrix}=\mathbf{B}$ as $\mathbf{Y}^{(k)} = \mathbf{Y}(t-\tau) + 0.5\tau (\mathbf{s}_1 + \mathbf{s}_2)$.
    \item To check the convergence, the solution $\mathbf{Y}^*$ of $\mathbf{A}\left(\mathbf{Y}^{(k)}\right)
    \begin{bmatrix}
        \mathbf{s}_1 \\
        \mathbf{s}_2
    \end{bmatrix}=\mathbf{B}$
    is compared with the solution from step 2: if the relative difference $\dfrac{\sum_{i=1}^N |Y_i^{(k)} - Y_i^{*}|}{\sum_{i=1}^N |Y_i^{*}|}$ is below the chosen tolerance $r_\mathrm{tol}$, $\mathbf{Y}^{(k)}$ is accepted as the solution $\mathbf{Y}(t)$. Otherwise, the steps 1-3 are repeated.
\end{enumerate}
Note that the evaluation of $\mathbf{A}\left(\mathbf{Y}^{(k)}\right)$ in step 3 is re-used at the next iteration or time step, so the convergence check does not introduce additional numerical costs associated with the evaluation of a transport model.
The choice of the parameter $\alpha\in (0,1]$ depends on the stiffness of an equation.

The convergence criterion in step 3 ensures that the accepted solution satisfies the original non-linear equation within the specified tolerance.
This rather strict criterion is chosen in order to avoid negating the advantages of the 2nd order accuracy of time stepping in the proposed numerical scheme.

\section{Results}
\label{sec:results}

In this section, the proposed numerical scheme is verified on a set of test examples in order to assess the spatial and temporal grids convergences, non-linear convergence and conservation properties.
The following shorthand notation is used throughout this section: iRK (implicit Runge-Kutta), BE (backward Euler) and IDO-CF (interpolated differential operation in conservative formulation described in Sec.~\ref{sec:spatial}).

\subsection{Spatial convergence}
\begin{figure}
    \centering
    \includegraphics[scale=0.8]{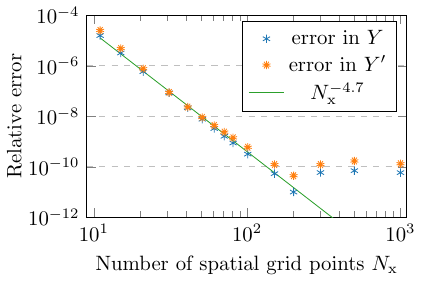}
    \caption{Convergence in spatial grid for IDO-CF scheme.}
    \label{fig:grid_convergence}
\end{figure}

    The convergence with the spatial grid size for IDO-CF is estimated by solving a linear ordinary differential equation
    \begin{equation}\nonumber
        \mathrm{e}^{1-x^2} Y + (-Y' + Y)'=\mathrm{e}^{1-x^2}\left(\mathrm{e}^{1-x^2} -4x^2 -2x +2 \right)
    \end{equation}
    with 
    the boundary conditions
    \begin{equation}
        Y'(x=0)=0; \quad Y(x=1)=1
    \end{equation}
    and the analytical solution $Y_\mathrm{true}(x)=\mathrm{e}^{1-x^2}$.
    The relative errors in the value and the derivative are estimated as
    \begin{equation}
        \dfrac{\sum_{i=1}^N |Y_i - Y_{\mathrm{true},i}|}{\sum_{i=1}^N |Y_{\mathrm{true},i}|}
        \quad\text{and}\quad
        \dfrac{\sum_{i=1}^N |Y'_i - Y'_{\mathrm{true},i}|}{\sum_{i=1}^N |Y'_{\mathrm{true},i}|}.
        \label{eq:errors_est}
    \end{equation}
    In Fig.~\ref{fig:grid_convergence} the relative errors are depicted as functions of the total number of spatial grid points $N_x$ for the proposed spatial discretisation scheme IDO-CF (Sec.~\ref{sec:spatial}).
    It demonstrates the convergence above the 4th order (the error is approximately proportional to $N_x^{-4.7}$) with the spatial grid size both in the value and in the derivative of the solution.
    The relative errors saturate at approximately $10^{-10}$ level due to finite precision arithmetics (see Sec. 4 in~\cite{Bell+etal2022}). 

\subsection{Time advance convergence}

\begin{figure}
    \centering
    \includegraphics[scale=0.8]{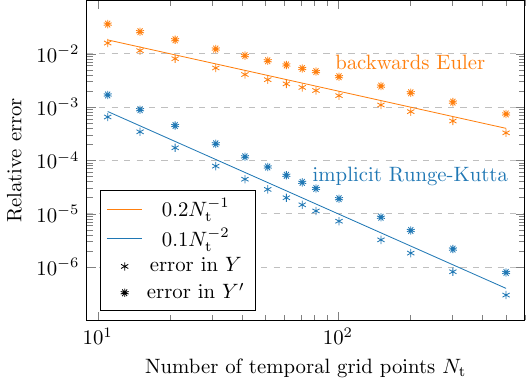}
    \caption{Convergence in temporal grid for IDO-CF scheme with BE (orange) and iRK (blue) time advance schemes.}
    \label{fig:time_convergence}
\end{figure}

To estimate the order of convergence with the time step size, a conservation equation for particles in cylindrical geometry~\cite{Hinton+Hazeltine1976} 
\begin{equation}
    x\frac{\partial Y}{\partial t} - (xY')' = 0
\end{equation}
is solved using IDO-CF spatial discretisation scheme both with iRK and with BE time advance schemes.
The analytical solution is
\begin{equation}
    Y_\mathrm{true}(x,t) = \frac{1}{4\pi t} \mathrm{e}^{-\frac{x^2}{4t}},
\end{equation}
and the boundary conditions are given by
\begin{equation}
    Y'(x=0,t)=0; \quad Y(x=5,t) = Y_\mathrm{true}(x=5,t).
\end{equation}
The temporal and spatial simulation domains are respectively $t\in [1,2]$ and $x\in [0,5]$, and the initial condition is $Y(x,t=1)=Y_\mathrm{true}(x,t=1)$.
The relative errors for the profile value and the derivative are estimated by Eq.~\eqref{eq:errors_est} at $t=2$ and are shown in Fig.~\ref{fig:time_convergence}.
As expected, the proposed IDO-CF scheme with iRK time advance demonstrates the second order convergence with the number of time steps $N_t$, while the IDO-CF with BE time advance~\eqref{eq:backwards_euler} has the first order convergence.
In both cases, the spatial grid with $N_x=101$ was used.
The error contribution due to spatial convergence is several orders of magnitude lower than the one from temporal convergence, and thus the time advance error dominates the total error.

\subsection{Nonlinear convergence}

\begin{figure}
    \centering
    \includegraphics[scale=0.8]{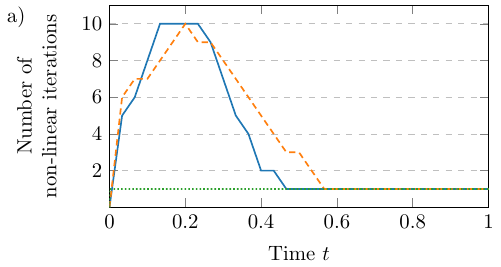}
    
    \includegraphics[scale=0.8]{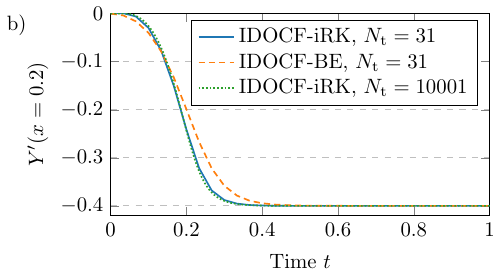}
    \caption{Nonlinear convergence for a stiff problem using IDO-CF spatial discretisation with iRK $N_\mathrm{t}=31$ (solid lines), with BE $N_\mathrm{t}=31$ (dashed lines), with iRK $N_\mathrm{t}=10001$ (dotted lines). (a) Number of non-linear iterations at every time step. (b) Profile derivative $Y'(x)$ at $x=0.2$ as a function of time. Total spatial grid points $N_\mathrm{x}=101$ for all three simulations.
    }
    \label{fig:nonlinear}
\end{figure}

\begin{figure}
    \centering
    \includegraphics[scale=1.0]{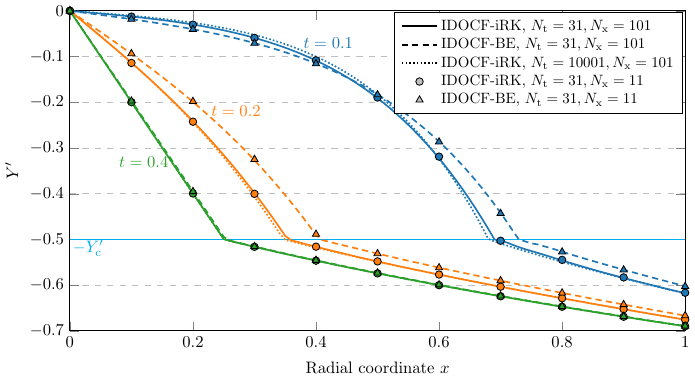}
    \caption{$Y'$ profiles for the stiff nonlinear problem at times $t=\{0.1,0.2,0.4\}$. Line styles correspond to the same simulations as in Fig.~\ref{fig:nonlinear}, and the profiles from reduced spatial grid simulations ($N_\mathrm{x}=11$) are shown with circles and triangles for iRK and BE respectively.
     }
    \label{fig:nonlinprofs}
\end{figure}

    To estimate the efficiency of the nonlinear convergence scheme using Picard-type operation with under-relaxation in combination with IDO-CF space discretisation, and iRK and BE time advance, the following nonlinear partial differential equation is solved
    \begin{equation}
        \frac{3}{2}\frac{\partial Y}{\partial t} - \frac{1}{x}
        \left( 
            xD(Y') Y'
        \right)'
        =
        4;
        \quad
        Y'(x=0)=0; \quad Y(x=1)=0,
        \label{eq:nonlin_problem}
    \end{equation}
    where the stiff transport model for diffusivity is used
    \begin{equation}\nonumber
        D(Y') =
        \begin{cases}
            D_0 + 10(|Y'|-Y'_c) & \text{for } |Y'|>Y'_c, \\
            D_0 & \text{for } |Y'|\leq Y'_c.
        \end{cases}
    \end{equation}
    Here, $D_0=1$ and $Y'_c=0.5$. At $t=0$, the initial condition is set to $Y=0$. 
    The time interval $t\in[0,1]$ is simulated with $N_t$ time steps. 
    The nonlinear convergence threshold $r_\mathrm{tol}=10^{-4}$ is set and Picard operation with under-relaxation with $\alpha=0.285$ is used in all the simulations. 
    Fig.~\ref{fig:nonlinear}a shows the number of nonlinear iterations (calls to the transport model) at each time step for three simulations using the IDO-CF numerical scheme with the iRK~\eqref{eq:lobatto1}-\eqref{eq:lobatto3} and BE~\eqref{eq:backwards_euler} time advance.
    The iRK with $N_t=31$ (solid line) and BE with $N_t=31$ (dashed line) require no more than 10 iterations at each time step to converge -- total amount of  iterations are 105 and 117 respectively.
    As a reference simulation with high resolution in time domain, IDO-CF with iRK $N_t=10001$ (dotted line) is used, which requires only one nonlinear iteration at each time step, that is, the total of 10001 iterations for the entire simulation.
    All three simulations are performed with $N_x=101$.
    The corresponding time traces of $Y'$ at the position $x=0.2$ are given in Fig.~\ref{fig:nonlinear}b.
    The iRK scheme with $N_t=31$ predicts the time evolution close to the reference simulation, while the BE scheme with $N_t=31$ has noticeable discrepancies.
    In Fig.~\ref{fig:nonlinprofs} the $Y'$ profiles are shown at $t=\{0.1,0.2,0.4\}$ for the three simulations described above. The critical gradient $-Y'_c$ is shown by the horizontal line in cyan colour.
    Additionally, an iRK simulation with $N_x=11, N_t=31$ (circle markers) shows a good agreement with the reference simulation, thus demonstrating the robustness of the numerical scheme even for the coarse spatial grids.

    The results in Figs.~\ref{fig:nonlinear}-\ref{fig:nonlinprofs} demonstrate that the proposed scheme, IDO-CF with iRK time advance and under-relaxation nonlinear convergence, allows for large time steps and for reduced spatial grids, while retaining high accuracy even for stiff non-linear problems. This in turn allows to significantly reduce the numerical costs associated with calls to the physical models.
    

\subsection{Conservation}

\begin{figure}
    \centering
    \includegraphics[scale = 1.0]{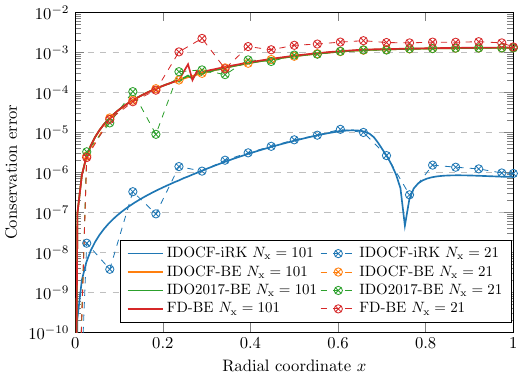}
    \caption{Conservation error evaluated with cubic splines for simulations with $N_t=161$ and $r_\mathrm{tol}=10^{-4}$ for a set of numerical schemes.}
    \label{fig:cons_x}
\end{figure}

    The nonlinear problem~\eqref{eq:nonlin_problem} was used to evaluate the conservation properties of the discretisation schemes.
    The normalised conservation error is estimated by integrating~\eqref{eq:nonlin_problem} in temporal and spatial coordinates
    \begin{align}
        \mathrm{Intertia}(x,t_1,t_2) = \frac{3}{2}\int_{0}^{x}\xi(Y(\xi,t_2)-Y(\xi,t_1))\mathrm{d}\xi, \label{eq:inertia} \\
        \mathrm{Flux}(x,t_1,t_2) = -\int_{t_1}^{t_2} \left[\xi D(Y'(\xi,t))Y'(\xi,t)\right]_{\xi=0}^{x}\mathrm{d}t, \label{eq:flux} \\
        \mathrm{Source}(x,t_1,t_2) = \int_{t_1}^{t_2} \int_0^x f \mathrm{d}x \mathrm{d}t = 2x^2(t_2-t_1), \\
        \mathrm{Error}(x,t_1,t_2) := \frac{\mathrm{Intertia}(x,t_1,t_2) + \mathrm{Flux}(x,t_1,t_2) - \mathrm{Source}(x,t_1,t_2)}{\mathrm{Source}(x_N,t_1,t_2)},
    \end{align}
    that is, the conservation error is the error in the total conserved quantity inside the flux surface $x$, normalised to the total integrated source.
    The profiles and diffusivities at the converged iterations are used in evaluating the conservation errors.
    The integration in Eqs.~\eqref{eq:inertia} and~\eqref{eq:flux} is performed using a cubic splines representation of integrands as a method not specific to the considered solvers for fair comparison.
    The time interval for evaluating the conservation is $[t_1,t_2]=[0.1,0.5]$.
    Fig.~\ref{fig:cons_x} compares 4 solvers: the proposed scheme IDO-CF with iRK (blue lines), IDO-CF with BE (orange lines), our implementation of the scheme of Ref.~\cite{Park+etal2017} which uses BE (green lines, IDO2017), and the 2nd order finite difference solver (FD) with BE (red lines).
    The solid lines are for $N_x=101$ and the dashed lines with circle-cross markers are for $N_x=21$.
    For all simulations $N_t=161$.
    The dominant contribution to the conservation error is due to the time integration scheme, and iRK provides approximately 2 orders of magnitude lower conservation error as compared to the BE schemes.
    Additionally, the proposed IDO-CF scheme gives exact conservation in each spatial cell, which is ensured by including the spatially integrated equation in the system~\eqref{eq:matrix}.
    FD and the scheme of~\cite{Park+etal2017} do not solve the spatially integrated equation exactly, and instead rely on the accuracy of the spatial discretisation scheme for conservation.
    For FD scheme this is demonstrated by the higher conservation error for $N_x=21$, which due to the error from the spatial accuracy exceeds the error due to the time accuracy.

\section{Conclusions}
\label{sec:conclusions}

This paper presents a numerical scheme for solving transport equations for tokamak plasmas.
The proposed numerical scheme has the exact conservation in spatial variable, above the 4th order of convergence in space, the 2nd order of convergence in time for the physics variable and its gradient, and a rapid non-linear convergence.
This enables robust simulations on reduced spatial and temporal grids, thus minimising the amount of calls to computationally expensive physical models.
The proposed scheme is compatible with non-equidistant grids.

The proposed scheme builds upon the existing numerical schemes of IDO family and provides the following advantages:
\begin{itemize}
    \item compared to~\cite{Imai+etal2008}, the implicit time advance is used, making the scheme suitable for stiff nonlinear equations;
    \item compared to~\cite{Imai+Aoki2006higher}, the conservative formulation in spatial discretisation is used, ensuring conservation of cell values;
    \item compared to~\cite{Park+etal2017}, which is specifically designed for plasma transport equations, our scheme has higher order of convergence in time and exact conservation in spatial coordinate. Additionally, the scheme proposed here does not require the second-order derivative of diffusivity and the first-order derivative of the source term, thus avoiding potential issues related to instabilities when taking numerical derivatives of transport coefficients computed by physical codes.
\end{itemize}

The considered numerical example with stiff transport model demonstrates that the second order time stepping method significantly improves the accuracy of the transient process modelling and the conservation properties in the simulation as compared to the first-order methods.

The proposed numerical scheme formulates the inputs in a generalised form suitable for integrated modelling codes.
Modification of the scheme to account for extra terms and coefficients in the generalised form of a transport equation is straightforward, as explained in the end of Sec.~\ref{sec:matrix}.
The extension of the spatial discretisation scheme IDO-CF to several spatial dimensions is described in Reference~\cite{Imai+etal2008}.
In the numerical examples we considered the diffusivity term as a source of a non-linearity, however, the non-linearities of other origin (such as coupling between transport equations) can be also taken into account inside the matrix $\mathbf{A}(\mathbf{Y})$ given in Sec.~\ref{sec:nonlin_converg}.

\appendix
 \section{Coefficients for the piece-wise polynomial approximation}
 \label{sec:polycoeff}
 
\newcommand{\dxr}{\delta_+}
\newcommand{\dxl}{\delta_-}
 
 With $\delta_+:=x_{i+1} - x_i$ and $\delta_- := x_{i-1} - x_i$ the matching conditions~\eqref{eq:matching1}-\eqref{eq:matching4} are rewritten as
  \begin{align}
        k_4^{(i)}\delta_+^4 + k_3^{(i)}\delta_+^3 + k_2^{(i)}\delta_+^2 + k_1^{(i)} \delta_+ + y_{i} &= y_{i+1}, \\
        k_4^{(i)}\delta_-^4 + k_3^{(i)}\delta_-^3 + k_2^{(i)}\delta_-^2 + k_1^{(i)} \delta_- + y_{i} &= y_{i-1}, \\
        k_4^{(i)}\frac{\delta_-^5}{5} + k_3^{(i)}\frac{\delta_-^4}{4} + k_2^{(i)}\frac{\delta_-^3}{3} + k_1^{(i)} \frac{\delta_-^2}{2} + y_{i}\delta_- &= -\tilde{y}_{i-1}, \\ 
        k_4^{(i)}\frac{\delta_+^5}{5} + k_3^{(i)}\frac{\delta_+^4}{4} + k_2^{(i)}\frac{\delta_+^3}{3} + k_1^{(i)} \frac{\delta_+^2}{2} + y_{i}\delta_+ &= \tilde{y}_{i}
    \end{align}
for each $i$th polynomial $P_i$, $i=2,..,N-1$.
This system of equations is linear with respect to $k_1^{(i)},k_2^{(i)},k_3^{(i)},k_4^{(i)}$ and the solution is
\begin{equation}
	\begin{split}
		k_{1}^{(i)} &= 
		- \frac{2 \dxr^{2}}{\dxl \left(\dxl -  \dxr \right)^2}
		y_{i-1}
		- 
		\left(\frac{4}{\dxr} + \frac{4}{\dxl}\right)
		y_{i}
		-
		\frac{2 \dxl^{2}}{\dxr \left(\dxl -  \dxr \right)^2}
		y_{i+1} \\
		&+
		\frac{\dxr^{2} \left(- 10 \dxl + 6 \dxr\right)}{\dxl^{2} \left(\dxl -  \dxr \right)^3}
		\tilde{y}_{i-1}
		+
		\frac{\dxl^{2}  \left(6 \dxl - 10 \dxr\right)}{\dxr^{2} \left(\dxl -  \dxr \right)^3}
		\tilde{y}_{i}
		= \mathbf{k}_{1}^{(i)}\mathbf{y},
	\end{split}
	\label{eq:k1}
\end{equation}

\begin{equation}
	\begin{split}
		k_2^{(i)} &=
		\frac{\dxr \left(6 \dxl + 3 \dxr\right)}{\dxl^{2}  \left(\dxl -  \dxr \right)^2}
		y_{i-1}
		+
		\left( \frac{3}{\dxr^{2}} + \frac{12}{\dxl \dxr} + \frac{3}{\dxl^{2}} \right)
		y_{i}
		+
		\frac{\dxl \left(3 \dxl + 6 \dxr\right)}{\dxr^{2} \left(\dxl -  \dxr \right)^2}
		y_{i+1} \\
		&+
		\frac{\dxr \left(30 \dxl^{2} - 6 \dxl \dxr - 6 \dxr^{2}\right)}{\dxl^{3} \left(\dxl -  \dxr \right)^3}
		\tilde{y}_{i-1}
		+
		\frac{\dxl \left(- 6 \dxl^{2} - 6 \dxl \dxr + 30 \dxr^{2}\right)}{\dxr^{3} \left(\dxl -  \dxr \right)^3}
		\tilde{y}_{i}
		= \mathbf{k}_2^{(i)}\mathbf{y},
	\end{split}
	\label{eq:k2}
\end{equation}

\begin{equation}
	\begin{split}
		k_3^{(i)} &=
		\frac{- 4 \dxl - 8 \dxr}{\dxl^{2} \left(\dxl -  \dxr \right)^2}
		y_{i-1}
		+
		\frac{8 \left(- \dxl - \dxr\right)}{\dxl^{2} \dxr^{2}}
		y_{i}
		+
		\frac{- 8 \dxl - 4 \dxr}{\dxr^{2} \left(\dxl -  \dxr \right)^2}
		y_{i+1} \\
		&+
		\frac{- 20 \dxl^{2} - 20 \dxl \dxr + 16 \dxr^{2}}{\dxl^{3} \left(\dxl -  \dxr \right)^3}
		\tilde{y}_{i-1}
		+
		\frac{16 \dxl^{2} - 20 \dxl \dxr - 20 \dxr^{2}}{\dxr^{3} \left(\dxl -  \dxr \right)^3}
		\tilde{y}_{i}
		= \mathbf{k}_3^{(i)}\mathbf{y},
	\end{split}
\end{equation}

\begin{equation}
	\begin{split}
		k_4^{(i)} &=
		\frac{5}{\dxl^{2} \left(\dxl -  \dxr \right)^2}
		y_{i-1}
		+
		\frac{5}{\dxl^{2} \dxr^{2}}
		y_{i}
		+
		\frac{5}{\dxr^{2} \left(\dxl -  \dxr \right)^2}
		y_{i+1} \\
		&+
		\frac{20 \dxl - 10 \dxr}{\dxl^{3} \left(\dxl -  \dxr \right)^3}
		\tilde{y}_{i-1}
		+
		\frac{- 10 \dxl + 20 \dxr}{\dxr^{3} \left(\dxl -  \dxr \right)^3}
		\tilde{y}_{i}
		= \mathbf{k}_4^{(i)}\mathbf{y}.
	\end{split}
\end{equation}
Note that $k_1^{(i)},k_2^{(i)},k_3^{(i)},k_4^{(i)}$ are linear combinations of $y_{i-1}, y_{i}, y_{i+1},\tilde{y}_{i-1}$ and $\tilde{y}_{i}$ and hence allow the representation $k_j^{(i)} = \mathbf{k}_j^{(i)}\mathbf{y}$.
The row vectors $\mathbf{k}_j^{(i)}$ only depend on the grid parameters $\delta_-,\delta_+$ and are non-zero only at the positions corresponding to the positions of $y_{i-1}, y_{i}, y_{i+1},\tilde{y}_{i-1},\tilde{y}_{i}$ in $\mathbf{y}$.
The IDO-CF spatial discretisation scheme uses the following quantities:
\begin{equation}
	y'_i = P'_i(\delta)\big|_{\delta=0} = k_1^{(i)} \quad \text{and} \quad y''_i = P''_i(\delta)\big|_{\delta=0} = 2 k_2^{(i)}, \quad i=2,..,N-1
	\label{eq:DY_DDY}
\end{equation}
for $\pmb{\ell}_i$ and $\tilde{\pmb{\ell}}_i$ in Eqs.~\eqref{eq:ell_i} and~\eqref{eq:tilde_ell_i}, and
\begin{equation}
	y'_{1} = P'_2(\delta)\big|_{\delta=\delta_-} = (4\mathbf{k}_4^{(2)}\delta_-^3 + 3\mathbf{k}_3^{(2)}\delta_-^2 + 2\mathbf{k}_2^{(2)}\delta_- + \mathbf{k}_1^{(2)}) \mathbf{y},
\end{equation}
\begin{equation}
	 y'_{N} = P'_{N-1}(\delta)\big|_{\delta=\delta_+} = (4\mathbf{k}_4^{(N-1)}\delta_+^3 + 3\mathbf{k}_3^{(N-1)}\delta_+^2 + 2\mathbf{k}_2^{(N-1)}\delta_+ + \mathbf{k}_1^{(N-1)}) \mathbf{y}
\end{equation}
for $\textbf{u}_1, \textbf{u}_N$ in Eq.\eqref{eq:u_l}.

\section{Linear transformations}
\label{sec:linear_transf}

The combination of Eqs.~\eqref{eq:ell_i},\eqref{eq:k1},\eqref{eq:k2} and~\eqref{eq:DY_DDY} provides the expression %for $\pmb{\ell}_i, i=2,..,N-1$
\begin{equation}
	\pmb{\ell}_i = d(x_i,t)2\mathbf{k}_2^{(i)} + (d'(x_i,t) - e(x_i,t))\mathbf{k}_1^{(i)} - e'(x_i,t) \mathbf{1}^{2N-1}_i, \quad i=2,..,N-1,
\end{equation}
where $\mathbf{1}^{2N-1}_i$ is a row vector of length $(2N-1)$ and the only non-zero element of value $1$ at the position $i$.
Similarly for Eq.~\eqref{eq:tilde_ell_i}
\begin{equation}
	\tilde{\pmb{\ell}}_i = 
	\begin{cases}
		d(x_i,t)\mathbf{k}_1^{(i)}  - e(x_i,t) \mathbf{1}^{2N-1}_{i}, \quad &i=2,..,N-1,\\
		d(x_i,t) (4\mathbf{k}_4^{(2)}\delta_-^3 + 3\mathbf{k}_3^{(2)}\delta_-^2 + 2\mathbf{k}_2^{(2)}\delta_- + \mathbf{k}_1^{(2)}) - e(x_i,t) \mathbf{1}^{2N-1}_{i}, \quad &i=1,
	\end{cases}
\end{equation}
and for Eq.~\eqref{eq:u_l}
\begin{equation}
	\mathbf{u}_i =
	\begin{cases}
		u_i \mathbf{1}^{2N-1}_{i} + v_i (4\mathbf{k}_4^{(2)}\delta_-^3 + 3\mathbf{k}_3^{(2)}\delta_-^2 + 2\mathbf{k}_2^{(2)}\delta_- + \mathbf{k}_1^{(2)}), \quad &i=1, \\
		u_i \mathbf{1}^{2N-1}_{i} + v_i (4\mathbf{k}_4^{(N-1)}\delta_+^3 + 3\mathbf{k}_3^{(N-1)}\delta_+^2 + 2\mathbf{k}_2^{(N-1)}\delta_+ + \mathbf{k}_1^{(N-1)}), &\quad i=N.
	\end{cases}
\end{equation}

\section*{Acknowledgements}

This work has received funding from the Swedish Research Council (2021-00182,``INFRAfusion'').

 \bibliographystyle{elsarticle-num} 
 \bibliography{refs}

\end{document}